\documentclass[onecolumn,showpacs,superscriptaddress,nobibnotes,nofootinbib,12pt]{revtex4}
\usepackage{amsmath,amssymb}
\usepackage{graphicx}

\begin{document}

\title{Comparing energy loss and $p_{\perp}$-broadening in perturbative QCD with strong coupling $\mathcal{N}=4$ SYM theory\footnote{This work is supported in part by the US Department of Energy.}}
\author{Fabio Dominguez}
\email{fad2111@columbia.edu}
\affiliation{Department of Physics, Columbia University, New York, NY, 10027, USA}
\author{C. Marquet\footnote{C.M. is supported by the European Commission under the FP6 program.}}
\email{cyrille@phys.columbia.edu}
\affiliation{Service de Physique Th\'eorique, CEA/Saclay, 91191 Gif-sur-Yvette cedex, France}
\affiliation{Department of Physics, Columbia University, New York, NY, 10027, USA}
\author{A. H. Mueller}
\email{amh@phys.columbia.edu}
\affiliation{Department of Physics, Columbia University, New York, NY, 10027, USA}

\author{Bin Wu\footnote{B.W. is supported by China Scholarship Council.}}
\email{bw2246@columbia.edu}
\affiliation{Department of Physics, Peking University, Beijing, 100871, P.R. China}
\affiliation{Department of Physics, Columbia University, New York, NY, 10027, USA}

\author{Bo-Wen Xiao}
\email{bowen@phys.columbia.edu}
\affiliation{Department of Physics, Columbia University, New York, NY, 10027, USA}

\begin{abstract}
We compare medium induced energy loss and $p_{\perp}$-broadening in perturbative QCD with that of the trailing string picture of SYM theory. We consider finite and infinite extent matter as well as relativistic heavy quarks which correspond to those being produced in the medium or external to it. When expressed in terms of the appropriate saturation momentum, we find identical parametric forms for energy loss in perturbative QCD and SYM theory. We find simple correspondences between $p_{\perp}$-broadening in QCD and in SYM theory although $p_{\perp}$-broadening is radiation dominated in SYM theory and multiple scattering dominated in perturbative QCD.
\end{abstract}

\date{\today}
\maketitle
\newpage
\section{Introduction}
The purpose of this paper is to compare energy loss and $p_{\perp}$-broadening for relativistic heavy quarks passing through a hot perturbative QCD plasma with that of the trailing string description\cite{Herzog:2006gh,Gubser:2006bz} for a hot SYM plasma. We shall see that for energy loss in infinite extent matter the formulas for perturbative QCD and for a strongly coupled $\mathcal{N}=4$ SYM theory are parametrically identical when expressed in terms of the relevant saturation momentum, $Q_{s}$. For $p_{\perp}$-broadening the SYM result is parametrically identical to a (subdominant) perturbative QCD term which in the strong coupling limit of SYM theory becomes the leading term. This comparison in infinite matter is then extended to finite extent matter where the trailing string picture has not yet been developed. We find results identical with infinite matter when expressed in terms of $Q_{s}$. We have not attempted comparison with a competitive theory of energy loss \cite{Liu:2006ug} based on a Wilson loop calculation because that approach is essentially perturbative with the SYM theory only being used to give an evaluation of the transport coefficient, $\hat{q}$.

In this introduction we motivate and summarize our results. In Sec.~\ref{compare}, these results are described in more detail with additional details and supporting evidence for our picture given in Sec.~\ref{distribution}, \ref{exact} and \ref{tensor}. 

We begin by recalling the picture of energy loss for a relativistic heavy quark\cite{Dokshitzer:2001zm} traveling through an infinite extent hot perturbative QCD plasma. We suppose the heavy quark to have rapidity $\eta$ at the time over which we focus. The heavy quark has a cloud of gluons, labeled by energy $\omega$ and transverse momentum $k_{\perp}$. Energy loss is dominated by the maximum energy gluons which are freed, or radiated, in the plasma\cite{Baier:1998kq,Zakharov:1996fv,Gyulassy:2000fs,Salgado:2003gb}. Gluons are freed if their transverse momentum is less than the saturation momentum corresponding to the gluon coherence time, $t_{c}$. This immediately gives
\begin{eqnarray}
-\frac{dE}{dt} \propto \alpha N_{c} \frac{\left(\omega\right)_{max}}{t_{c}}=\alpha N_{c} Q_{s}^{2}, \label{heavy}
\end{eqnarray}
where we have used $t_{c}=\frac{\left(\omega\right)_{max}}{Q_{s}^2}$ as the relevant coherence time. The condition $\frac{\left(\omega\right)_{max}}{Q_{s}}=\cosh \eta$ as well as $Q_{s}^2=\hat{q} t_{c}$ converts Eq.~(\ref{heavy}) to a more standard form. The $\alpha N_{c}$ in Eq.~(\ref{heavy}) is just the probability of having a gluon, in the kinematic regime in question, in the heavy quark wavefunction.

We now note that Eq.~(\ref{heavy}) is also true for a strong coupling $\mathcal{N}=4$ SYM plasma if the replacement $\alpha N_{c} \rightarrow \frac{\sqrt{\lambda}}{2 \pi}$ is made, where $\lambda = g_{YM}^2 N_{c}$, and if the saturation momentum is now taken as $Q_{s}^2 \propto \left(t_{c}T^2\right)^2$ as found in Ref.~\cite{Hatta:2007cs}. (The stronger dependence of $Q_{s}^2$ on $t_c$ in SYM theory, as compared to perturbative QCD, comes about because high energy scattering in SYM theory is dominated by a $J \simeq 2$ singularity rather than the $J \simeq 1$ singularity in perturbative QCD.) At first sight it may seem strange that a perturbative picture, giving Eq.~(\ref{heavy}), could also lead to a correct result for a strongly coupled $\mathcal{N}=4$ SYM theory. However, we know that for a static heavy quark in the vacuum the distribution of energy in the fields surrounding the center of the heavy quark is the same for strong coupling SYM theory and for QCD, or even QED\cite{Chesler:2007sv,Gubser:2007nd}. As we show in Sec.~\ref{distribution}, this leads to an energy distribution in momentum space for a high energy heavy quark in SYM theory exactly as in QCD when one replaces $\alpha N_{c}$ by $\frac{\sqrt{\lambda}}{2\pi}$. The saturation momentum then determines the scale at which quanta in the heavy quark wavefunction are freed in the plasma, and this is the only dynamics necessary to get the parametric form Eq.~(\ref{heavy}).

Further evidence for this picture comes from $p_{\perp}$-broadening of a heavy quark in infinite extent matter. In QCD this broadening occurs mainly by direct multiple scattering of the heavy quark by thermal quanta\cite{Bodwin:1981fv,Baier:1996sk}. There is a subdominant effect, down by $\alpha N_{c}$, where the heavy quark gets increased transverse momentum by radiation of gluons. However, in strong coupling SYM theory $\alpha N_{c}$ is replaced by $\frac{\sqrt{\lambda}}{2\pi}$ and now radiation becomes dominant. Thus by simply allowing the emissions which give the energy loss to have a random transverse momentum, whose size is fixed to be that of the saturation momentum, one arrives at a dominant form for $p_{\perp}$-broadening  in $\mathcal{N}=4$ SYM theory (see Eq.~(\ref{ptb5})) which is the same as that for radiative broadening in QCD and which agrees with calculations done using the trailing string picture. 

Calculations of small oscillations on the trailing string\cite{Gubser:2006nz,CasalderreySolana:2007qw} lead one to identifying a point on the string which corresponds to a horizon on the metric induced on the worldsheet of the trailing string as it moves through the QCD plasma. This horizon corresponds to a $u-$value (see the metric Eq.~(\ref{metric1}))  equal to $u_{s}=\pi T \sqrt{\cosh \eta}$ which is the same as the value of the saturation momentum for the heavy quark in the plasma. In our picture all quanta in the heavy quark having transverse momenta less than $Q_s$ are freed while those having transverse momenta greater than $Q_{s}$ remain part of the heavy quark. In Sec.~\ref{tensor}, we verify that the part of the trailing string having $u>u_{s}$ corresponds to energy which spatially lies very close to the center of the heavy quark while the $u<u_{s}$ part of the trailing string corresponds to energy lagging too far behind the heavy quark center to be considered part of it. This part of the trailing string corresponds to waves in the plasma, and this matches our interpretation in terms of quanta which have been freed from the wavefunction of the heavy quark. We also note that $u_{s}$  is the same point as that identified in Ref.~\cite{Hatta:2008tx} as separating branching in the vacuum, $u>u_{s}$, from branching in the medium, $u<u_{s}$. In the trailing string picture this agrees with our interpretation of the $u>u_{s}$ part of the string as being the same as that of a heavy quark in the vacuum while the $u<u_{s}$ part of the string is interpreted as quanta separate from the heavy quark. 

Energy loss in finite matter is more subtle in both QCD and in $\mathcal{N}=4$ SYM theory. The first issue that arises is whether a heavy quark going through a finite extent of matter is "bare" or "dressed"\cite{Baier:1998kq,Zakharov:1996fv}. If the heavy quark does not have too large a rapidity, $\cosh \eta \ll \left(LT\right)^2$ in a SYM plasma, then the matter is effectively infinite extent and we are back to our earlier discussion. However, if $\cosh \eta$ is large there is a strong distinction between bare and dressed heavy quarks. A dressed heavy quark is one that has its gluon cloud completely developed as it enters the medium. In the trailing string picture a vertical string going from $u=0$ to $u=u_m$ entering the medium traveling at rapidity $\eta$ corresponds to a dressed quark entering the medium. For a dressed heavy quark energy loss simply corresponds to freeing those quanta having transverse momentum less than $Q_{s}$, with the length determining $Q_{s}$ being given by the length $L$ of the matter. In the case of the trailing string it, parametrically, corresponds to freeing all the string lying below $u=u_{s}$. The result, 
Eq.~(\ref{eloss6}), takes the same form for a QCD plasma and for a SYM plasma. 

For bare quarks we are faced with a more serious challenge. In QCD one can simply produce a heavy quark-antiquark pair in a hard collision and then allow, say, the heavy quark to go through the medium. In SYM theory consider a heavy quark-antiquark pair initially at rest and very close together. We then rapidly accelerate the quark and antiquark in opposite directions using an external electric field. When the quark has reached the desired rapidity we turn off the electric field and let the quark continue to pass through the medium. We have found an exact solution for the motion of the string during the period of acceleration, and this solution is described in Sec.~\ref{exact} and in Ref.~\cite{pt}. One can see from this solution, and the discussion in Sec.~\ref{compare}, that now the heavy quark is missing much of its cloud of quanta as it passes through the medium. The key to determining medium induced energy loss is an understanding of the time at which components of the final dressed heavy quark are formed. In QCD this is well understood. In Sec.~\ref{compare} we motivate, but ultimately conjecture that the formation times are the same for strong coupling SYM theory and perturbative QCD. This leads to an energy loss formula as in Eq.~(\ref{heavy}) for QCD and the same, with $\alpha N_{c} \rightarrow \frac{\sqrt{\lambda}}{2 \pi}$, for SYM theory but where now $Q_{s}^2=\hat{q} L$ in perturbative QCD and $Q_{s}^2 \sim \left(L T^2\right)^2$ for hot SYM matter. We note in passing that the induced energy loss for hot SYM matter goes as the cube of the length of the material in contrast to a quadratic law in QCD\cite{Baier:1996kr} while the total $p_{\perp}^2$ picked up traversing the medium goes like $L^2$ for hot SYM matter instead of the linear dependence on $L$ in hot perturbative QCD\cite{Bodwin:1981fv,Baier:1996sk}.

\section{Comparison between QCD and SYM results for energy loss and $
p_{\perp}$-broadening}\label{compare}

In this section we give a detailed comparison of results on energy loss and $
p_{\perp}$-broadening for a heavy quark in QCD with corresponding results in
$\mathcal{N}=4$ SYM theory. Although the results for the strong coupling SYM theory
have been derived (In some cases the derivations will be given for the first
time in later section of this paper.) using the AdS/CFT correspondence\cite{Maldacena:1997re,Witten:1998qj,Gubser:1998bc} with
the SYM theory, we here emphasize the comparison of QCD directly with the
SYM theory. We rely on explicit calculations on the string theory side of
the correspondence to give confirmation of the basic validity of our
picture. We start with energy loss in infinite hot matter where both the QCD
and SYM theory results are well known. What we are here noting is that when
expressed in terms of the relevant saturation momentum, the results for QCD
and SYM are essentially identical for energy loss, however, the physical
picture for $p_{\perp}$-broadening differs in these two theories.

\subsection{QCD energy loss of heavy quarks in infinite extent hot matter}

Consider a heavy quark of mass $M$ passing through hot matter of temperature
$T$. We suppose $T/M \ll 1$. Suppose at some time the heavy quark is moving
at rapidity $\eta$ so that its energy is $M \cosh \eta$. We also assume, for
reasons to be explained below, that $\cosh \eta \ll M^{3}/\hat{q} $ with the
$\hat{q}$ the transport coefficient of the plasma.

The dominant cause of energy loss of the heavy quark is gluon radiation\cite{Dokshitzer:2001zm,Baier:1998kq,Zakharov:1996fv,Gyulassy:2000fs,Salgado:2003gb}
induced by the medium, at least when $\cosh \eta \gg 1$ which is the region
with which we shall be concerned. Induced gluon radiation is caused by
gluons in the wavefunction of the heavy quark being freed by interaction with
the medium. Suppose a gluon in the quark's wavefunction has energy $\omega$
and transverse momentum $k_{\perp}$. Such a gluon will be radiated if $
k_{\perp}\le Q_{s}$ with $Q_{s}$ the saturation momentum of the medium
corresponding to a length given by the coherence time, $t_{c}$, of the
gluon. The energy loss will be dominated by gluons having $k_{\perp}$, and $
\omega$, as large as possible. Thus, setting $k_{\perp}=Q_{s}$, we have
\begin{eqnarray}
\hat{q} t_{c}=Q_{s}^{2} \label{tqs}
\end{eqnarray}
and using
\begin{eqnarray}
t_{c}=\frac{\omega}{Q_{s}^{2}},  \label{tc}
\end{eqnarray}
gives
\begin{eqnarray}
Q_{s}^{4}=\hat{q} \omega.  \label{qhat}
\end{eqnarray}
Gluons having $\frac{\omega}{k_{\perp}}=\frac{\omega}{Q_{s}}> \cosh \eta$
are strongly suppressed in the heavy quark wavefunction\cite{Dokshitzer:2001zm}. Taking $\omega =
Q_{s} \cosh \eta$ will give the maximum energy and, combined with Eq.~(\ref
{qhat}), gives
\begin{eqnarray}
Q_{s}^{3}=\hat{q} \cosh \eta  \label{qs}
\end{eqnarray}
as the relevant saturation momentum. Now we see why it is necessary to keep $
\cosh \eta \ll M^{3}/\hat{q} $, because this is the condition that $\frac{
k_{\perp}^2}{M^2}=\frac{Q_{s}^{2}}{M^2}\ll 1$, which condition is the
essential requirement separating heavy quark from light quark radiation
dynamics. Now it is a straightforward task to give a parametric form of the
rate of energy loss of a heavy quark passing through a plasma as
\begin{eqnarray}
-\frac{dE}{dt} \propto \alpha_{s}N_{c}\frac{\omega}{t_{c}}.  \label{eloss1}
\end{eqnarray}
The $\alpha_{s}N_{c}$ in Eq.~(\ref{eloss1}) corresponds to the number of
gluons, in the relevant kinematic domain, in the quark wavefunction while $t_c$ corresponds to the time over which the gluon energy, $\omega$, is
emitted. Using Eq.~(\ref{tc}) gives
\begin{eqnarray}
-\frac{dE}{dt} \propto \alpha_{s}N_{c} Q_{s}^{2}.  \label{eloss2}
\end{eqnarray}
where, $Q_{s}$ should be evaluated using Eq.~(\ref{qs}).

The difference between heavy and light quarks lies not in changing Eq.~(\ref
{eloss2}), which remains valid for light quarks, but in Eq.~(\ref{qs}) which
for light quarks energy loss becomes
\begin{eqnarray}
Q_{s}^{2} \simeq \sqrt{E \hat{q} }  \label{qslight}
\end{eqnarray}
with $E$ the energy of the light quark.

\subsection{Energy loss of a heavy quark in SYM theory in infinite extent
hot matter}

The problem of energy loss for heavy quarks in a $\mathcal{N}=4$ SYM plasma has been
well studied\cite{Herzog:2006gh,Gubser:2006bz}. Using the AdS/CFT correspondence allows one to evaluate the
energy loss in terms of the energy flowing toward the horizon through a
string which goes from a $D7$ brane, whose position is given by the heavy
quark mass, toward the horizon of the metric. The metric in the $AdS_5$
space can be written as
\begin{equation}
ds^{2}=R^{2}u^{2}\left[ -f\left(u\right)dt^{2}+ d\vec{x}
^{2} \right]+\frac{du^{2} R^2}{u^{2} f\left(u\right)},
\label{metric1}
\end{equation}
with
\begin{equation}
f\left(u\right) = 1- \left(\frac{u_{h}}{u}\right)^{4}
\end{equation}
and where $u_{h}=\pi T$ and $T$ is the temperature of the plasma. The $D7$
brane is located at a position $u_m$ which is related to the heavy quark's
mass by
\begin{equation}
M=\frac{\sqrt{\lambda}u_{m}}{2\pi}  \label{mass}
\end{equation}
corresponding to a string at rest in the vacuum having length $u_m$, as it
falls straight down from $u_m$ to $u=0$ in the fifth dimension. The energy
density of the string is
\begin{equation}
\frac{dE}{du}=\frac{\sqrt{\lambda}}{2\pi}.  \label{e-d}
\end{equation}
In Eqs.~(\ref{mass}) and (\ref{e-d}), $\lambda$ is given by $\lambda =
g_{YM}^{2} N_{c}$. In the energy loss problem one keeps the string moving at
a constant rapidity $\eta$ where one imagines an external constant
"electric" field acting on the end of the string on the $D7$ brane and
furnishing the force necessary to keep the string moving at constant
velocity $v$ where $\cosh \eta = 1/\sqrt{1-v^2}$. The resulting rate of
energy loss, that is the rate of work done by the external electric field,
is\cite{Herzog:2006gh,Gubser:2006bz}
\begin{equation}
-\frac{dE}{dt}=\frac{\pi \sqrt{\lambda}}{2}T^2 v^2 \cosh \eta \simeq \frac{%
\pi \sqrt{\lambda}}{2}T^2 \cosh \eta  \label{eloss3}
\end{equation}
where we have only consider the case $\cosh \eta \gg 1$.

We are now going to show that Eq.~(\ref{eloss3}) agrees with Eq.~(\ref%
{eloss2}) if one uses the saturation momentum appropriate to the $\mathcal{N}=4$ SYM
plasma instead of Eq.~(\ref{qs}). To get the replacement of Eq.~(\ref{qs})
for the SYM plasma, we recall that the saturation momentum corresponding to
a length $L$ of such a plasma is\cite{Hatta:2007cs}
\begin{equation}
Q_{s}\left(L\right) \sim L T^2.  \label{qsl}
\end{equation}
Using $t_c$ in Eq.~(\ref{tc}) as the relevant length , just as we used for
calculating the energy loss in QCD, and taking $\omega/Q_{s} = \cosh \eta$,
one gets
\begin{eqnarray}
Q_{s}^2 \sim T^2 \cosh \eta  \label{qseta}
\end{eqnarray}
from Eq.~(\ref{qsl}). Using Eq.~(\ref{qseta}) in Eq.~(\ref{eloss3}), one
arrives at
\begin{eqnarray}
-\frac{dE}{dt} \sim \sqrt{\lambda} Q_{s}^2.  \label{eloss4}
\end{eqnarray}
Except for the replacement of $\alpha N_{c}$ by $\sqrt{\lambda}$
Eq.~(\ref{eloss2}) and Eq.~(\ref{eloss4}) are the same. The replacement of $
\alpha N_{c}$ by $\sqrt{\lambda}$ is to be expected and already occurs in
comparing the static energy density, the electric field squared, of a heavy
quark in QCD with that of a heavy quark in $\mathcal{N}=4$ SYM theory. As we show in
Sec.~\ref{distribution}, this naturally leads a gluon distribution in a heavy quark to be the
same in QCD and in $\mathcal{N}=4$ SYM theory except for the replacement of $\alpha
N_{c}$ by $\sqrt{\lambda}$. Our picture of energy loss of a heavy quark in
an infinite plasma is thus essentially identical in QCD and in $\mathcal{N}=4$ SYM. We
do note, however, that we do not have control of constant factors in our
description of energy loss in a $\mathcal{N}=4$ SYM plasma.

\subsection{Transverse momentum broadening of a heavy quark in a QCD plasma}

The picture of transverse momentum broadening of a high energy quark in QCD
is very simple and is the same for light quarks as for heavy quarks\cite{Bodwin:1981fv,Baier:1996sk}. As the
quark passes through the plasma, it interacts with quanta of the plasma
through multiple single gluon exchange, each exchange giving a random
transverse momentum to the quark, and two gluon exchanges, necessary to keep
probability conservation. The formula
\begin{equation}
\frac{dp^{2}_{\perp}}{dt} = \hat{q}.  \label{ptb1}
\end{equation}
gives the rate of increase in the typical $p^{2}_{\perp}$ that the quark
picks up in passing through the medium in terms of the transport coefficient
of medium. One can also write this relation as
\begin{equation}
\frac{dp^{2}_{\perp}}{dt} = \frac{dQ_{s}^{2}\left(t\right)}{dt}.
\label{ptb2}
\end{equation}
where $Q_{s}\left(t\right)$ is the saturation momentum of a length, $t$, of
the medium where we do not distinguish between time and length intervals for
our relativistic heavy quark. The result given in Eq.~(\ref{ptb1}) and Eq.~(%
\ref{ptb2}) comes completely from random multiple scattering of the quark in
the medium. There is also a contribution from emission of gluons stimulated
by the medium. The same gluon emissions which give the energy loss indicated
in Eq.~(\ref{eloss2}) naturally give a $p_{\perp}$-broadening
\begin{equation}
\left(\frac{dp^{2}_{\perp}}{dt}\right)_{\text{radiation}} \sim \alpha_{s}
N_{c} \frac{dQ_{s}^{2}\left(t\right)}{dt}.  \label{ptb3}
\end{equation}
Eq.~(\ref{ptb3}) is perhaps obvious once one writes Eq.~(\ref{eloss2}) in
the form
\begin{equation}
-\frac{dE}{dt} \sim \alpha_{s} N_{c} \frac{d \omega \left(t\right)}{dt}
\label{eloss5}
\end{equation}
where $\omega \left(t\right)$ is given by Eq.~(\ref{qhat}) with the $t$
-dependence of $Q_s$ given by Eq.~(\ref{tqs}), identifying $t_c$ with $t$.
Then Eq.~(\ref{ptb3}) and Eq.~(\ref{eloss5}) directly give the energy loss
and transverse momentum broadening due to gluon radiation stimulated by the
medium. The contribution of Eq.~(\ref{ptb3}) is usually neglected because it
is parametrically smaller than the contribution given in Eq.~(\ref{ptb2})
from multiple scattering. However, as we shall see below in a SYM plasma,
transverse momentum broadening of a heavy quark is dominated by gluon
radiation because $\alpha N_{c}$ will be replaced by the large parameter, $%
\sqrt{\lambda}$.

\subsection{Transverse momentum broadening of a heavy quark in an infinite
extent SYM plasma}

The calculation of transverse momentum broadening in a SYM plasma is done in
terms of fluctuations on the corresponding string on the gravity side of the
AdS/CFT correspondence. The result is\cite{Gubser:2006nz,CasalderreySolana:2007qw}
\begin{equation}
\frac{dp^{2}_{\perp}}{dt}= 2\pi \sqrt{\lambda} T^{3} \sqrt{\cosh \eta}.
\label{ptb4}
\end{equation}
If we identify the saturation momentum $Q_s$ as in Eq.~(\ref{qseta}), then
one can write
\begin{equation}
\frac{dp^{2}_{\perp}}{dt} \sim \sqrt{\lambda} T^{2} Q_{s}.  \label{ptb5}
\end{equation}
or, alternatively, as
\begin{equation}
\frac{dp^{2}_{\perp}}{dt} \sim \sqrt{\lambda} \frac{dQ_{s}^{2}\left(t\right)%
}{dt}.  \label{ptb6}
\end{equation}
where, again, we identify $t$ with $t_c$ the coherence time of the gluons
whose emission dominates both the energy loss and transverse momentum
broadening of the heavy quark. Eq.~(\ref{ptb6}) is identical, after $\alpha
N_{c} \leftrightarrow \sqrt{\lambda} $, with Eq.~(\ref{ptb3}) adding
confirmation to our picture of energy loss and $p_{\perp}$-broadening as due
to gluon emission in the SYM plasma. Once one writes formulas in terms of
the relevant saturation momentum, the SYM and QCD pictures become identical.
Of course, in the case of $p_{\perp}$-broadening what is a subdominant
effect in QCD, gluon emission, becomes dominant in the SYM plasma.

\subsection{The saturation momentum and the trailing string}

Our picture of infinite extent plasmas has been that a heavy quark loses
energy and gets transverse momentum broadening by the emission of gluons
having transverse momentum equal to the saturation momentum of the medium
corresponding to a medium length given by the coherence length of the
emitted gluons which carry the same rapidity as the parent heavy quark. Now
we are going to identify this picture more closely with the trailing string
picture.

In the trailing string picture, a heavy quark moving at constant rapidity
through a SYM plasma correponds to a string moving at the same rapidity in the background metric, Eq.~(\ref{metric1}) on the AdS side of the correspondence.
The shape of the string is\cite{Herzog:2006gh,Gubser:2006bz}
\begin{equation}
z\left( u\right) =z_{0}+vt+\frac{v}{2u_{h}}\left[ \frac{\pi }{2}-\tan
^{-1}\left( \frac{u}{u_{h}}\right) -\coth ^{-1}\left( \frac{u}{u_{h}}\right) 
\right]   \label{shape}
\end{equation}%
where
\begin{equation}
u_{h}<u<u_{m}=\frac{2\pi M}{\sqrt{\lambda }}  \label{range}
\end{equation}%
with $M$ the mass of the heavy quark. In our picture, the parts of the
string having $u_{m}>u>Q_{s}$ correspond to gluons actually in the heavy
quark while those parts of the string having $u_{h}<u<Q_{s}$ correspond to
freed matter which is no longer part of the heavy quark. The flow of energy
past the point
\begin{equation}
Q_{s}=u_{s}\equiv \pi T\sqrt{\cosh \eta },  \label{qs2}
\end{equation}
with $Q_{s}$ given as in Eq.~(\ref{qseta}), corresponds to the freeing of
gluons from the heavy quark and into QCD plasma. Let us now see that $u\sim
u_{s}$ is a natural point on the string for separating the heavy quark from
waves in the plasma. Let $\Delta z\left( u\right) $ be the distance between
a point on the trailing string and a corresponding point on a string, having
the same motion at $u=u_{m}$, moving in the vacuum. Using Eq.~(\ref{shape})
for $u$ large, one easily finds
\begin{equation}
\Delta z\left( u\right) =-\frac{vu_{h}^{2}}{3}\left( \frac{1}{u^{3}}-\frac{1
}{u_{m}^{3}}\right) .  \label{dz1}
\end{equation}
Now we expect that the range of z-values allowed in a heavy quark moving at
velocity $v$ in the vacuum is given by
\begin{equation}
\delta z\left( u\right) \simeq -\frac{1}{u}\frac{1}{\cosh \eta },
\label{dz2}
\end{equation}
where the first factor on the right hand side of Eq.~(\ref{dz2}) is the
infrared-ultraviolet correspondence while the second factor is due to
Lorentz contraction. Those region of $u$ having $\Delta z\left( u\right)
\leq \delta z\left( u\right) $ can naturally be part of the heavy quark
while those region of $u$ having $\Delta z\left( u\right) >\delta z\left(
u\right) $ lag too far behind the core of the heavy quark to be part of it.
Comparing Eq.~(\ref{dz1}) and Eq.~(\ref{dz2}), we see that the separation
point, up to constant factor of order one, is given by Eq.~(\ref{qs2}). In
Sec.~\ref{tensor}, we give a more complete discussion of this issue where, instead of
relying on Eq.~(\ref{dz2}), we compare the four-dimensional energy-momentum
tensor for a trailing string with that of the string corresponding to a
heavy quark moving in the vacuum and reach the same conclusion as above.

Another argument for Eq.~(\ref{qs2}) as the natural separation point between
what belongs to the heavy quark and what has been freed into the medium
comes from the calculation of transverse momentum broadening from small
fluctuation on the trailing string. In the differential equation governing
the small fluctuations $u=u_{s}$ is a regular singular point which
corresponds to a horizon of the metric on the worldsheet of the string\cite{Gubser:2006nz,CasalderreySolana:2007qw}.
Waves on the worldsheet of the string going from large $u$ can disappear
into the horizon but one requires that waves not come out of the horizon
toward large values of $u$.\cite{Herzog:2002pc,Son:2007vk}

\subsection{Energy loss and $p_{\perp }$-broadening for finite matter and
for dressed quarks}

In dealing with hot matter of finite length in QCD, the energy
loss problem is very different for dressed and bare quarks. For
heavy ion collisions where high transverse momentum quarks are
produced in the collision as bare quarks, without the accompanying
gluon cloud which characterize a dressed quark, the dressed quark
energy loss problem is of little interest. Nevertheless we begin
with this problem in our discussion of finite matter because it is
relatively straightforward.

The situation is as follows: A high energy heavy quark prepared at
early times impinges on a target of hot matter, say a cube, of length
$L$. We wish to determine how much energy is lost and how much
transverse momentum broadening occurs as the heavy quark passes
through the medium. If the rapidity of the heavy quark is such
that $\cosh \eta \ll \left(LT\right)^2$(For QCD the condition is
$\cosh \eta \ll \sqrt{\hat{q}L^3}$), then the coherence time of
all the gluons in the heavy quark is much less than $L$ and the
matter is effectively of infinite extent. Thus, we consider here only the opposite case where $\cosh \eta \gg \left(LT\right)^2$. We also suppose $\frac{M}{LT^2}\gg 1$ so that a small fraction of the heavy quark's energy is lost as it passes through the matter. 

When $\cosh \eta \gg \left(LT\right)^2$, the picture is essentially the same in QCD and for a SYM 
plasma. Gluons in the heavy quark wavefunction having $k_{\perp}<Q_{s} \sim LT^2$ will be freed in passing through the matter while those having $k_{\perp} \geq Q_{s}  $ are not freed. Since the gluons which dominate the energy loss have $\omega \simeq k_{\perp} \cosh \eta$, the energy loss will be 
\begin{eqnarray}
-\frac{dE}{dt} \propto \binom{\alpha N_{c}}{\sqrt{\lambda}} \frac{Q_{s}\cosh \eta}{L}.  \label{eloss6}
\end{eqnarray}
with $\alpha N_{c}$ referring to QCD and $\sqrt{\lambda}$ to the SYM case. For SYM the result Eq.~(\ref{eloss6}) agrees with Eq.~(\ref{eloss3}), the infinite matter result, when Eq.~(\ref{qsl}) is used for $Q_{s}$. For QCD the result Eq.~(\ref{eloss6}) differs from the case of infinite matter, and we remark that we have not subtracted the "factorization term"\cite{Gyulassy:1993hr,Baier:1994bd}. We emphasize, however, that when written in terms of the saturation momentum the QCD and SYM results are of the same form once one makes the $\alpha N_{c} \leftrightarrow \sqrt{\lambda}$ identification in passing between theories. 

For transverse momentum broadening the formulas are 
\begin{eqnarray}
\frac{dp_{\perp}^2}{dt} \propto \binom{1}{\sqrt{\lambda}} \frac{Q_{s}^2}{L}.  \label{ptb7}
\end{eqnarray}
where again the upper value is for QCD while the lower value is for a SYM plasma. The QCD result is the same as Eq.~(\ref{ptb1}) for infinite matter while the SYM result takes the same form as infinite matter, as given in  Eq.~(\ref{ptb5}), but now one must use Eq.~(\ref{qsl}) for $Q_{s}^2$. The argumentation leading to Eq.~(\ref{ptb7}) is identical to that giving Eq.~(\ref{eloss6})  and, again, here in the SYM plasma transverse momentum broadening is radiation dominated. 

\subsection{Energy loss and $p_{\perp}$-broadening for bare heavy quarks in finite extent matter}

The quarks, whether light or heavy, that are produced in hard collisions in a relativistic heavy ion collision are initially bare, that is they have been produced without the gluon cloud that accompanies a quark which is part of a high energy hadron. While the energy loss of bare quarks has been widely discussed in the QCD literature\cite{Baier:1998kq,Zakharov:1996fv,Gyulassy:2000fs,Salgado:2003gb}, it has so far not been treated in the literature on the SYM plasma. On the string theory side of the AdS/CFT correspondence, in order to create a heavy quark which is initially bare it is necessary to consider a heavy quark-antiquark pair initially with little or no separation and then rapidly accelerate the quark and the antiquark in opposite direction by, say, an electric field $E_{f}$ which is constant in space and time and points along the $z$-axis. After a rapid acceleration over a time $t_{1}$, the electric field can be turned off and one has a quark-antiquark pair with large relative energy in a situation similar to that occurring in a hard collision\cite{Herzog:2006gh}. We have found an exact solution which satisfies the equations coming from the Nambu-Goto action for the quark-antiquark pair accelerating in the vacuum. This will be discussed in much more detail in Sec.~\ref{exact} and in Ref.~\cite{pt}, while here we give the solution and use it to outline how to get energy loss and $p_{\perp}$-broadening in a SYM plasma. 

The solution, in the vacuum, described above is 
\begin{eqnarray}
z=\pm \sqrt{t^{2}+\frac{c^{2}}{u_{m}^2}-\frac{1}{u^{2}}}  \label{solution0}
\end{eqnarray}
where the external electric field is 
\begin{eqnarray}
E_{f}=\frac{2\pi M^2}{\sqrt{\lambda }c}=\frac{\sqrt{\lambda} u_{m}^2}{2\pi c}\label{ef}
\end{eqnarray}
The quark is at $z>0$ and antiquark is at $z<0$. $M$ is the mass of the heavy quark as usual. The constant $c$ characterizes the strength of the electric field and $c\geq 1$ is required, as is clear from Eq.~(\ref{solution0}). We imagine $c$ to be large but we also suppose $\frac{u_m}{c}$ is very large compared to the temperature of the medium which we shall shortly introduce. At a time $t_{1}$, we shall turn off the electric field beyond which time Eq.~(\ref{solution0}) no longer applies. It is straightforward to see that, for $t<t_{1}$, the part of the string in the region  $u_{m}/c<u<u_{m}$ moves with a rapidity $\eta\left(t\right)$ given by
\begin{eqnarray}
\cosh \eta =\frac{1}{\sqrt{1-v^2}} \simeq \frac{u_{m}t}{c}. \label{cosheta}
\end{eqnarray}
We now suppose that 
\begin{eqnarray}
 \frac{u_{m}t}{c}\gg 1, \label{cosheta2}
\end{eqnarray}
however, we also imagine that $t_{1}$ is a small time compared to the inverse temperature of the medium. By choosing $u_m$ sufficiently large there is no difficulty in having $c\gg 1$, $u_{m} t_{1}/c \gg 1$, and $t_{1}T \ll 1$ satisfied. 

Before considering motion in a plasma, let us examine a little further the vacuum evolution of our heavy quark-antiquark system. When  $c/u_{m}<t<t_{1}$, it is useful to consider the $z>0$ part of the string as two parts. Part A consists of the $u_{m}/c \leq u \leq u_{m}$ region of the string, while part B is the $\frac{1}{\sqrt{t^2+c^2/u_{m}^2}}\leq u < u_{m}/c$ region of the string. Much as in the trailing string , part A is part of the heavy quark, while part B corresponds to radiated energy. Indeed with this interpretation we can calculate the power radiated at time $t$ as 
\begin{eqnarray}
P=\frac{d}{dt}\left[ E_{f}t -\cosh \eta \left(t\right)\frac{\sqrt{\lambda}}{2\pi}\left(u_{m}-u_{m}/c\right)\right]. \label{power}
\end{eqnarray}
The first term on the right hand side of Eq.~(\ref{power}) gives the rate at which the electric field puts energy into the system. Noting that the energy density of a string at rest is $\frac{dE}{du}=\frac{\sqrt{\lambda}}{2\pi}$, we see that the second term is the rate of growth of the energy in the "straight section" of the string in the region $u_{m}/c \leq u \leq u_{m}$. Eq.~(\ref{power}) gives 
\begin{eqnarray}
P_{\text{radiated}}=\frac{\sqrt{\lambda}}{2\pi} \frac{E_{f}^2}{M^2}. \label{power2}
\end{eqnarray}
We note that the answer for a classical electron accelerating in a constant electric field is\cite{jackson} 
\begin{eqnarray}
P_{\text{electron}}=\frac{2 e^2}{3} \frac{E_{f}^2}{M^2}, \label{power3}
\end{eqnarray}
parametrically the same as Eq.~(\ref{power2}) with the replacement $e^2 \leftrightarrow \sqrt{\lambda}$. 

Armed with our understanding that the part of the string $u< u_{m}/c$ is emitted radiation and no longer part of the heavy quark, we are now ready to insert the medium. We suppose that the acceleration during $0<t<t_{1}$ takes place at one end of a hot SYM plasma whose $z$-extent is $L$. At $t=t_{1}$, we turn off the electric field and let our heavy quark pass through the length $L$ of the plasma. There are two separate cases to consider: (i), when $\cosh \eta \left(t_{1}\right)\equiv \cosh \eta \ll \left(LT\right)^2$ and (ii), when $\cosh \eta \gg \left(LT\right)^2$. 

In case(i), the situation is just like that of a trailing string in an infinite medium. In vacuum the separation point between that part of the string corresponding to the heavy quark and previously radiated energy is at $u=\frac{u_{m}}{c}\frac{t_{1}}{t}=\frac{\cosh \eta}{t}$  when $t> t_{1}$. At a time $t=\frac{\sqrt{\cosh \eta}}{T}$  this point crosses the separation point, $u=T \sqrt{\cosh \eta}$, of the trailing string at which point the heavy quark energy loss and transverse momentum broadening becomes that of the trailing string. This time at which the string becomes a trailing string is much less than $L$ so that the whole problem becomes that of an infinite medium trailing string problem. In case (i), $\cosh \eta$ is small enough that the system quickly adjust to the trailing string scenario. 

\begin{figure}
\begin{center}
\includegraphics[width=8cm]{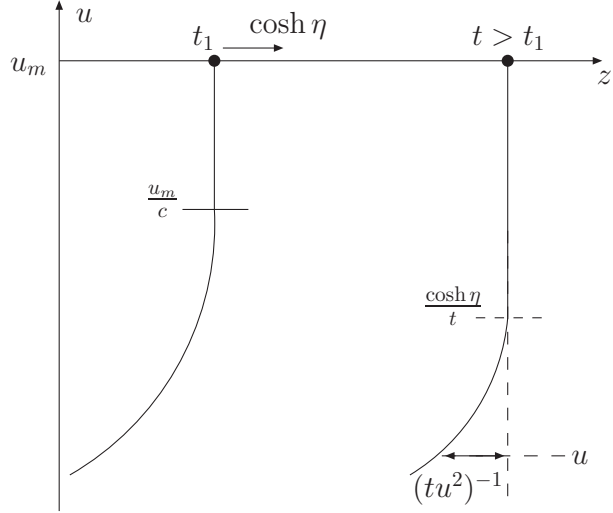}
\end{center}
\caption[*]{The accelerating string in the vacuum. At the time $t_1,$ when the electric field is turned off, the quark rapidity is $\cosh \eta \left(t_{1}\right)\equiv \cosh \eta,$ and the part of the string with $u<u_m/c$ is emitted radiation. At a later time t, the heavy quark is rebuilding its gluon cloud, and for $u\ll\cosh\eta/t,$ only soft components with $k_z< t u^2$
contribute to the heavy quark. When put in a medium of length L (such that $(LT)^2\ll\cosh\eta$), radiated gluons must also have $u\leq Q_s,$ therefore the hardest, dominant, radiations have $k_z=LQ_s^2.$}
\label{acstring}
\end{figure}

In case(ii) the situation is much more subtle. To better appreciate the issues, let us first suppose the motion of the evolving string to take place in the vacuum. Then, when the electric field is turned off at $t=t_{1}$, the system will go toward a final configuration of a heavy quark moving in the vacuum at constant velocity along with some radiation. The radiation all goes toward $u=0$ as explained in detail in Ref.~\cite{Hatta:2008tx}. For us the key issue is at what time the quanta making up the heavy quark are formed. We know that the separation point, which is at $u=u_{m}/c$ for $t<t_{1}$, moves down in $u$ as 
\begin{eqnarray}
\left(u\right)_{\text{separation}}=\frac{\cosh \eta}{t}, \label{sepa}
\end{eqnarray}
when $t>t_{1}$, essentially according to free branching. However, once the electric field has been turned off this separation point need no longer be a perfect separation between what is part of the heavy quark and what is radiated energy. Although almost all the energy in the region $u<\left(u\right)_{\text{separation}}$ lags too far behind the core of the heavy quark to be part of it, at a distance
\begin{eqnarray}
\Delta z\left(u,t\right) \sim \frac{1}{t u^2} \label{dz3}
\end{eqnarray}
behind the core as is clear from Eq.~(\ref{solution0}), the components of the heavy quark having $k_{\perp} \sim u$ and $k_{z} \sim t u^2 \ll u \cosh \eta$ spatially overlap with the evolving string as is evident in the discussion in Sec.~\ref{tensor} and on Fig.\ref{acstring}. If  these components of the evolving string do contribute to the final heavy quark, they are formed at a time $t_c$ where
\begin{eqnarray}
t_{c} \sim \frac{k_{z}}{u^2}\sim \frac{k_{z}}{k_{\perp}^2}, \label{tc3}
\end{eqnarray}
which matches our expectation of the time at which such components are naturally formed. Of course the components of the heavy quark carrying most of the energy, having $k_{z}\simeq u \cosh \eta$, would still be formed much later at the time when the separation point, given in Eq.~(\ref{sepa}), reaches u. 

It is the identification of $\frac{k_{z}}{k_{\perp}^2}\simeq \frac{k_{z}}{u^2}$ with the time at which soft components, those where $\frac{k_{z}}{u \cosh \eta} \ll 1$, of the free heavy quark are formed which drives us toward identifying a (small) part of the evolving string having $u<\left(u\right)_{\text{separation}}$ as contributing to the heavy quark. We now make this conjecture, namely that after the electric field has been turned off at $t=t_{1}$, the parts of the heavy quark having $k_{\perp} \sim u$  and $k_{z}$ are formed at a time $t_{c} \sim k_{z}/u^2$
out of a small fraction, a fraction $\frac{k_{z}}{u \cosh \eta}$, of the energy of the evolving string located at $u$ at time on the order of $t_c$. 

Now let us return to the string evolving in the SYM medium. After the electric field is turned off, the system goes toward radiation and a partially formed heavy quark passing through the medium. In our present situation, $\cosh \eta \gg \left(LT\right)^2$, the heavy quark is only fully formed after leaving the medium. Now medium induced energy loss is determined by which components of the heavy quark which are forming in the medium are inhibited from doing so by the medium. Clearly all momentum components $k_{\perp} \sim u <Q_{s}$ will be inhibited. Thus the energy loss will be equal to that part of the free heavy quark string having $u<Q_{s}$ and $\frac{k_{z}}{u^2}< L$, or 
\begin{eqnarray}
-\frac{dE}{dt} \sim \sqrt{\lambda} \frac{\left(k_{z}\right)_{max}}{L}\sim \sqrt{\lambda} Q_{s}^2, \label{elossf}
\end{eqnarray}
exactly as in Eq.~(\ref{eloss4}), but where now
\begin{eqnarray}
Q_{s} \sim L T^2. \label{qsl2}
\end{eqnarray}
Similarly the transverse momentum broadening is given as 
\begin{eqnarray}
\frac{dp_{\perp}^2}{dt} \sim \sqrt{\lambda} \frac{d Q_{s}^2}{dL}, \label{ptbf}
\end{eqnarray}
as in Eq.~(\ref{ptb6}) but now with Eq.~(\ref{qsl2}) again determining $Q_{s}$. 

The rather elaborate argument we have just given for energy loss, and which also give Eq.~(\ref{qsl2}) for $p_{\perp}$-broadening is essentially identical to the understanding of energy loss in perturbative QCD in a finite length medium\cite{Baier:1998kq,Zakharov:1996fv,Gyulassy:2000fs,Salgado:2003gb,Baier:1996sk,Baier:1996kr}. So it should be no surprise that there the result is
\begin{eqnarray}
-\frac{dE}{dt} = \frac{\alpha N_{c}}{4}\hat{q} L=\frac{\alpha N_{c}}{4} Q_{s}^2, \label{elossfp}
\end{eqnarray} 
and 
\begin{eqnarray}
\frac{dp_{\perp}^2}{dt} = \hat{q}=  \frac{d Q_{s}^2}{dL}, \label{ptbfp}
\end{eqnarray}
when $\cosh \eta > \sqrt{\hat{q} L^3}$ and where the $Q_{s}^2$ in Eqs.~(\ref{elossfp}) and (\ref{ptbfp}) is given by $Q_{s}^2=\hat{q}L$. 

\section{Lienard-Wiechert calculation of gluon distribution in heavy quark.}\label{distribution}
It is known that the energy-momentum tensor of a heavy particle at rest in SYM theory has the same form as the one obtained from classical electrodynamics (except for a normalization factor)\cite{Chesler:2007sv,Gubser:2007nd}. This statement is easily generalized for a heavy particle moving at constant velocity since the energy-momentum tensor is related to the previous case by a Lorentz boost.

In this section we will show that, for a relativistic heavy particle, the classical result for the energy density in momentum space agrees with the quantum perturbative calculation to lowest order. For this purpose we will compute the energy density in momentum space by Fourier transforming the Lienard-Wiechert potential of a moving charge at constant velocity and compare this result to the distribution of photons in the wavefunction of a fast moving charge.

For a particle moving at constant velocity along the z-axis, the potential is given by
\begin{align}
A_{0}(x)&=\frac{e}{4\pi \sqrt{(z-vt)^{2}+(1-v^{2})x_{\bot}^{2}}},  \\
A_{z}(x)&=v A_{0}(x).
\end{align}
Taking a three-dimensional Fourier transform
\begin{eqnarray}
A_{0}(\vec{k},t)= \frac{e}{(2\pi)^{3/2}}\frac{e^{-i vtk_{z}}}{k_{\bot}^{2}+(1-v^{2})k_{z}^{2}}.
\end{eqnarray}
The corresponding fields are
\begin{align}
\vec{E}(\vec{k},t)&=-\frac{i e^{-ivtk_{z}}}{(2\pi)^{3/2}} \frac{e(\vec{k}_{\bot},(1-v^{2})k_{z})}{k_{\bot}^{2}+(1-v^{2})k_{z}^{2}} \\
\vec{B}(\vec{k},t)&=i\vec{k}\times \vec{A}(\vec{k},t),
\end{align}
and the energy density
\begin{equation}
\frac{d\mathcal{E}}{d^{2}k_{\bot}dk_{z}} = \frac{e^{2}}{2(2\pi)^{3}}\frac{(1+v^{2})k_{\bot}^{2}+(1-v^{2})^{2}k_{z}^{2}}{(k_{\bot}^{2}+(1-v^{2})k_{z}^{2})^{2}}.
\end{equation}
For $v$ close to 1 we take
\begin{equation} \label{endens}
\frac{d\mathcal{E}}{d^{2}k_{\bot}dk_{z}} = \frac{e^{2}}{(2\pi)^{3}}\frac{k_{\bot}^{2}}{(k_{\bot}^{2}+(1-v^{2})k_{z}^{2})^{2}}.
\end{equation} \\

On the other hand, the wavefunction of a fast moving charge can be calculated perturbativelly to first order in the coupling
\begin{equation}
|\psi_{p}\rangle = |p\rangle + \sum_{\lambda}\int d^{3}k \; \psi_{\lambda}(\vec{k}) |p-k;k,\lambda\rangle
\end{equation}
where
\begin{equation}
\psi_{\lambda}(\vec{k}) = \frac{1}{E_{\vec{p}-\vec{k}}+E_{\vec{k}}-E_{\vec{p}}} \langle p-k;k,\lambda | H_{I} |p\rangle
\end{equation}
and $H_{I}$ is the interaction Hamiltonian. Assuming $p_{z}\gg m,k_{z}\gg k_{\bot}$ we find
\begin{equation}
\psi_{\lambda}(\vec{k}) = \frac{k_{z}}{k_{\bot}^{2}+(1-v^{2})k_{z}^{2}} \frac{e}{(2\pi)^{3/2}} \frac{1}{\sqrt{E_{\vec{p}}\, E_{\vec{p}-\vec{k}}\, |\vec{k}|}} p^{\mu} \epsilon_{\mu}^{(\lambda)}(\vec{k})^{*}.
\end{equation}
In Coulomb gauge, the polarization vectors can be written as
\begin{equation}
\epsilon_{\mu}^{(\lambda)}(\vec{k}) = (0,\vec{\epsilon}_{\bot}^{(\lambda)},-\frac{\vec{k}_{\bot}\cdot \vec{\epsilon}_{\bot}^{(\lambda)}}{k_{z}}),
\end{equation}
so, when $v$ is close to 1, we find
\begin{equation}
\psi_{\lambda}(\vec{k}) = -\frac{e}{(2\pi)^{3/2}\sqrt{|\vec{k}|}} \frac{\vec{k}_{\bot}\cdot \vec{\epsilon}_{\bot}^{(\lambda)}(\vec{k})^{*}}{k_{\bot}^{2}+(1-v^{2})k_{z}^{2}}.
\end{equation}

From this distribution we can extract the energy density in momentum space corresponding to the photons in the wavefunction.
\begin{align}
\frac{d\mathcal{E}}{d^{2}k_{\bot}dk_{z}} &= \sum_{\lambda} |\vec{k}| |\psi_{\lambda}(\vec{k})|^{2} \\
&= \frac{e^{2}}{(2\pi)^{3}}\frac{k_{\bot}^{2}}{(k_{\bot}^{2}+(1-v^{2})k_{z}^{2})^{2}}.
\end{align}
This is the same as \eqref{endens}, so we can conclude that the two pictures give the same result. Even though this calculation was done in the abelian case, the QCD calculation is analogous and gives the same result. Because the form of the energy momentum tensor in classical electrodynamics and strong coupling SYM theory is the same, the above calculation shows that when the strong classical fields of a high energy heavy quark in SYM theory are resolved into quanta, by requiring that the classical energy in a given wave number mode be given by the number of quanta in that mode times the mode frequency, the result is the same as in lowest order perturbation theory. Thus, despite the strong coupling of our SYM theory the distribution of numbers of quanta and energy in the various wave number modes is just the same as in lowest order perturbation theory up to a normalizing constant. Once we appreciate that the distribution of energy and momentum in modes is essentially the same as in lowest order perturbation it is no longer so surprising that the formulas for $dE/dt$ and $dp_\perp^2/dt$ have the same form as in perturbative QCD when expressed in terms of $Q_s$ since it is $Q_s$ which determines which modes are freed when the heavy quark passes through matter.

\section{The accelerating string}\label{exact}

\subsection{The accelerating string solution}

We set up our accelerating string calculation as follows: a
quark-antiquark pair is imbedded in a brane located at $u=u_{m}$, and a net
electric field $E_{f}$ is imposed in the brane which accelerates the quark and
antiquark at a constant acceleration in their own proper frame (An
additional small electric field $E_{f2}$ which balances the attractive force between the quark and
antiquark is also understood.).

The metric of the resulting vacuum $Ad_5$ space can be
written as
\begin{eqnarray}
ds^{2} &=&R^{2}\left[ \frac{du^{2}}{u^{2}}-u^{2}dt^{2}+u^{2}\left(
dx^{2}+dy^{2}+dz^{2}\right) \right] \\
&=&\frac{R^{2}}{\bar{u}^{2}}\left( d\bar{u}^{2}-dt^{2}+dx^{2}+dy^{2}+dz^{2}\right) ,
\end{eqnarray}%
where $R$ is the curvature radius of the $AdS_5$ space and $\bar{u}=\frac{1}{u}$. The dynamics of a classical string is characterized
by the Nambo-Goto action,
\begin{equation}
S=-T_{0}\int d\tau d\sigma \sqrt{-\det g_{ab}}
\end{equation}%
where $\left( \tau ,\sigma \right) $ are the string world-sheet coordinates,
and $-\det g_{ab}=-g$ is the determinant of the induced metric. $T_{0}$ is
the string tension. We define $X^{\mu }\left( \tau ,\sigma \right) $ as a
map from the string world-sheet to the five dimensional space time, and
introduce the following notation for derivatives: $\dot{X}^{\mu }=\partial
_{\tau }X^{\mu }$ and $X^{\prime \mu }=\partial _{\sigma }X^{\mu }$. When
one chooses a static gauge by setting $\left( \tau ,\sigma \right) =\left(
t,u\right) $, and defines $X^{\mu }=\left( t,u,x\left( t,u\right)
,0,0\right) $, it is straightforward to find that
\begin{eqnarray}
-\det g_{ab} &=&\left( \dot{X}^{\mu }X_{\mu }^{\prime }\right) ^{2}-\left(
\dot{X}^{\mu }\dot{X}_{\mu }\right) \left( X^{\prime \mu }X_{\mu }^{\prime
}\right)  \label{decayA} \\
&=&R^{4}\left( 1-\dot{x}^{2}+u^{4}x^{\prime 2}\right) .
\end{eqnarray}%
Therefore, the equation of motion of the classical string reads:
\begin{equation}
\frac{\partial }{\partial u}\left( \frac{u^{4}x^{\prime }}{\sqrt{-g}}\right)
-\frac{\partial }{\partial t}\left( \frac{\dot{x}}{\sqrt{-g}}\right)=0
\end{equation}

\begin{figure}
\begin{center}
\includegraphics[width=12cm]{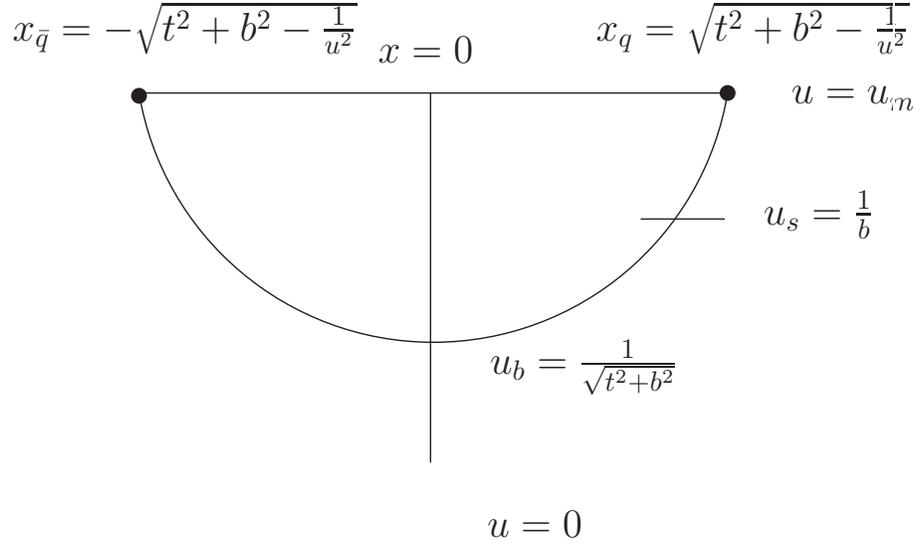}
\end{center}
\caption[*]{Illustrating the accelerating string.}
\label{accstring}
\end{figure}

In general, this equation is a non-linear differential equation which
involves two variables and two derivatives. Thus it is notoriously hard to
solve directly when $x\left( t,u\right) $ is a non-trivial function of $\left(
t,u\right) $. Fortunately, we have been able to find an exact solution which corresponds to
the accelerating string. The solution reads,
\begin{equation}
x=\pm \sqrt{t^{2}+b^{2}-\frac{1}{u^{2}}}  \label{solution}
\end{equation}
where the $+$ part represents the right moving part of the string and the $-$
part yields the left moving part of the string, together with the smooth
connection in the middle (see Fig.~\ref{accstring}). The quark and antiquark are accelerating and moving 
away from each other. The constant $b$ can be fixed by the boundary
condition, and corresponds to $b=\frac{c}{u_{m}}$ from Sec.~\ref{compare}, where $c$ is a dimensionless quantity. It is very easy to check that Eq.~(\ref{solution}) satisfies the
equation of motion by noting that $\sqrt{-g/R^4}=\frac{b}{\sqrt{t^{2}+b^{2}-
\frac{1}{u^{2}}}}$.

Following Herzog et al \cite{Herzog:2006gh} , one can compute the canonical
momentum densities associated with the accelerating string,
\begin{eqnarray}
\pi _{\mu }^{0} &=&-T_{0}\frac{\left( \dot{X}^{\nu }X_{\nu }^{\prime
}\right) X_{\mu }^{\prime }-\left( X^{\prime \nu }X_{\nu }^{\prime }\right)
\dot{X}_{\mu }}{\sqrt{-g}}, \\
\pi _{\mu }^{1} &=&-T_{0}\frac{\left( \dot{X}^{\nu }X_{\nu }^{\prime
}\right) \dot{X}_{\mu }-\left( \dot{X}^{\nu }\dot{X}_{\nu }\right) X_{\mu
}^{\prime }}{\sqrt{-g}}.
\end{eqnarray}%
The energy density is given by $\pi _{t}^{0}$,
\begin{equation}
\frac{dE}{du}=-\pi _{t}^{0}=\frac{T_{0}R^{4}}{\sqrt{-g}}\left(
1+u^{4}x^{\prime 2}\right) .
\end{equation}%
Thus the total energy of the right half string at time $t$ is,
\begin{equation}
\int_{u_{b}}^{u_{m}}\frac{dE}{du}du=\frac{T_{0}R^{2}u_{m}}{b}\sqrt{
t^{2}+b^{2}-\frac{1}{u_{m}^{2}}}.
\end{equation}
Moreover, the energy flow is given by $\pi _{t}^{1}$,
\begin{equation}
\frac{dE}{dt}=\pi _{t}^{1}=\frac{T_{0}R^{4}}{\sqrt{-g}}u^{4}x^{\prime }\dot{x
}.
\end{equation}
Thus the net energy\footnote{The total energy being put into the system should be the sum of the work done by $E_{f}$ and
$E_{f2}$. $E_f$ is the field giving constant acceleration in the absense of Coulomb attraction between the heavy quark and antiquark. The field $E_{f2}$ is included to cancel that Coulomb force and is therefore time dependent. However only $E_{f}$ contributes to the non-Coulomb net energy increase and to the constant acceleration $E_{f}/M$.} being put into the right half string from $0$ to $t$ is,
\begin{equation}
\left. \int_{0}^{t}\frac{dE}{dt}dt\right\vert _{u=u_{m}}=\frac{
T_{0}R^{2}u_{m}}{b}\left( \sqrt{t^{2}+b^{2}-\frac{1}{u_{m}^{2}}}-\sqrt{b^{2}-
\frac{1}{u_{m}^{2}}}\right) ,
\end{equation}
with the second term in the bracket being the initial energy deposited in
the string. Also $b^{2}-\frac{1}{u_{m}^{2}}\geq 0$ is assumed
for consistency. Therefore, from energy conservation, one can
easily fix the constant $b$ by setting $E_{f}=\frac{T_{0}R^{2}u_{m}}{b}$,
then,
\begin{equation}
b=\frac{M}{E_{f}}=\frac{\sqrt{\lambda}u_{m}}{2\pi E_{f}},
\end{equation}
where $M=T_{0}R^{2}u_{m}$ is the mass of the heavy quark and $T_{0}R^2=\frac{\sqrt{\lambda}}{2\pi}$ according to the AdS/CFT correspondence. It is now very easy to see the physical interpretation of the constant $b$ as the reciprocal of the constant acceleration $a$, i.e., $a=\frac{E_{f}}{M}=\frac{1}{b}$. 

In addition, although $\frac{\partial x}{\partial t}=\frac{t}{\sqrt{%
t^{2}+b^{2}-\frac{1}{u^{2}}}}$ exceeds $1$ when $u$ becomes smaller
than $1/b$, one can compute the speed at which energy travels by the following,
\begin{equation}
v=\frac{\partial x}{\partial t}+\frac{\partial x}{\partial u}\frac{du}{dt}=
\frac{t}{t^{2}+b^{2}}\sqrt{t^{2}+b^{2}-\frac{1}{u^{2}}}, \label{velocity}
\end{equation}
and find that $v\leq 1$ at all times. In arriving at the above result, one needs to
look at the hypersurface where energy is constant. Then, one can also obtain $
\frac{du}{dt}=-\frac{\partial E}{\partial t}/\frac{\partial E}{\partial u}=-
\frac{ut}{t^{2}+b^{2}}$. Finally, the Lorentz boost factor of the string
reads,%
\begin{equation}
\cosh \eta =\frac{1}{\sqrt{1-v^{2}}}=\frac{t^{2}+b^{2}}{\sqrt{\left(
t^{2}+b^{2}\right) b^{2}+\frac{t^{2}}{u^{2}}}},
\end{equation}%
and it reduces to $\frac{t}{b}=\frac{t u_{m}}{c}$ in the large $t$ and $u$ limits.

\subsection{Energy loss due to radiation}

In a following paper\cite{pt}, we will explicitly show that there is a scale $
u_{s}=\frac{1}{b}$ separating the soft part(the lower part) of the
string from the hard part of the string(the upper part). The upper part, which
moves together with the heavy quark, corresponds to the co-moving hard partons in the
heavy quark wave function; The lower part ($u<u_{s}$) of the string, which
lies far behind the heavy quark, is emitted radiation, and it is no longer
part of the heavy quark.

Therefore, the radiated energy at time $t$ is
\begin{equation}
E_{\text{radiation}}=\int_{u_{b}}^{u_{s}}\frac{dE}{du}=\frac{\sqrt{\lambda }}{2\pi }\frac{t
}{b^{2}},
\end{equation}
and the radiation power reads, 
\begin{equation}
P=\frac{dE_{\text{radiation}}}{dt}=\frac{\sqrt{\lambda }}{2\pi }\frac{E_{f}^{2}}{M^{2}}, \label{powerc}
\end{equation}
in agreement with Eq.~(\ref{power}). By using the same picture, we can also estimate the $p_{T}$ and $p_{L}$
broadening due to radiation. At large time limit, one finds
\begin{equation}
\frac{dp_{T}^{2}}{dt}\propto \frac{\sqrt{\lambda }}{2\pi }\frac{u_{s}^{2}}{t}%
\sim \frac{\sqrt{\lambda }}{2\pi }\frac{1}{b^{2}t},
\end{equation}%
where $\sqrt{\lambda }$ basically counts the number of partons being
emitted, $u_{s}^{2}$ is the typical momentum square of the emitted partons,
and $t$ is just the time scale of the system. Similarly, one finds
\begin{equation}
\frac{dp_{L}^{2}}{dt}\propto \frac{\sqrt{\lambda }}{2\pi }\frac{\omega
_{s}^{2}}{t}\sim \frac{\sqrt{\lambda }}{2\pi }\frac{t}{b^{4}},
\end{equation}%
with $\omega _{s}\sim \frac{1}{\Delta x}\simeq u_{s}^{2}t$ being the typical
energy of the emitted partons and $\Delta x$ being the longitudinal
separation between the quark and the string at $u=u_{s}$. Moreover, after identifying $u_s$ with $Q_{s}$, one discovers that the
coherence time $t=\frac{\omega}{u_{s}^2}$ in this accelerating string scenario coincides with the one in QCD (see Eq.~(\ref{tc})). 

The exact evaluation by employing random fluctuation analysis will be provided in Ref.~\cite{pt}, and it yields
\begin{eqnarray}
\frac{dp_{T}^{2}\left( t\right) }{dt} &=&\frac{\sqrt{\lambda }}{\pi ^{2}}
\frac{1}{b^{2}\sqrt{t^{2}+b^{2}}}=\frac{\sqrt{\lambda }}{\pi ^{2}}
\frac{a^3}{\sqrt{a^2t^{2}+1}}, \\
\frac{dp_{L}^{2}\left( t\right) }{dt} &=&\frac{\sqrt{\lambda }}{2\pi ^{2}}
\frac{\sqrt{t^{2}+b^{2}}}{b^{4}}=\frac{\sqrt{\lambda }}{2\pi ^{2}}a^3\sqrt{a^2t^{2}+1}.
\end{eqnarray}
Finally, we have checked that our solution Eq.~(\ref{powerc}) gives the same radiation as that of Mikhailov's general radiation formulas\cite{Mikhailov:2003er} when specialized to the case of constant acceleration.

\subsection{Accelerating string in non-zero temperature $AdS_{5}$ space}

The metric of the resulting AdS black brane solution in 5 dimension with
finite temperature $T$ can be written as
\begin{eqnarray}
ds^{2}=R^{2}\left[ \frac{du^{2}}{u^{2}\left( 1-\frac{u_{h}^{4}}{u^{4}}%
\right) }-u^{2}\left( 1-\frac{u_{h}^{4}}{u^{4}}\right) dt^{2}+u^{2}\left(
dx^{2}+dy^{2}+dz^{2}\right) \right] ,
\end{eqnarray}
with $u_{h}=\pi T$. Unfortunately, we are unable to find an exact
accelerating string solution in this case. Nevertheless, we can give some
semi-quantitative discussion in the $a=1/b=\frac{E_{f}}{M}\gg \pi T$ limit. 

Suppose the quark-antiquark pair now is put in the non-zero temperature $AdS_{5}$ spacetime, and the electric field which accelerates the heavy quarks is imposed from $t=0$ to $t=\infty$, we expect at  time $t\simeq \frac{1}{b}\frac{1}{\left(\pi T\right)^2}$ that the string starts to lose a significant amount of energy into the plasma. In the string system, energy is flowing into the system from the top of the string, while it is being absorbed by the plasma from the bottom part of the string. As time increases, the acceleration slows down and energy loss to the plasma increases. As a result, the string trajectory starts to evolve towards the trailing string solution from the accelerating string solution. Ultimately, the energy loss per unit time will be the same as the energy gained per unit time from the electric field. Thus, the trailing string trajectory is fully formed and a final velocity is then reached. Therefore, one gets
\begin{eqnarray}
\left(\frac{dE}{dt}\right)_{\text{loss}}+\left(\frac{dE}{dt}\right)_{\text{gain}}=0. \label{balance}
\end{eqnarray}
Using Eq.~(\ref{eloss3}) for energy loss per unit time and $E_{f}$ for energy gain per unit time, one gets the final $\cosh \eta_{f}$:
\begin{eqnarray}
\cosh \eta_{f}\simeq\frac{2\pi E_{f}}{\sqrt{\lambda}\left(\pi T\right)^2} \label{finalg}
\end{eqnarray}
In arriving above formula, we have assumed that $\cosh \eta_{f} \gg 1$. $\cosh \eta_{f}$ is independent of $u_{m}$ as a result of energy balance between the external electric field and dissipative plasma. In the end, we can estimate the typical time which the string needs to reach the trailing string solution from Eq.~(\ref{finalg}) and Eq.~(\ref{cosheta}), and obtain,
\begin{eqnarray}
t_{f}\sim \frac{2\pi M}{\sqrt{\lambda}\left(\pi T\right)^2}. \label{ftime}
\end{eqnarray}

\section{Energy momentum tensor in 4-dimension from trailing and accelerating strings}\label{tensor}

In this section, we will give a quantitative confirmation of our
picture by investigating $T_{\mu \nu}(t,\vec{x})$, the
four-dimensional energy-momentum tensor of the "gluonic" field
around a heavy quark. In the large $N_c$ limit, $T_{\mu
\nu}(t,\vec{x})$ can be obtained from $h_{MN}$, the small metric
perturbations to the $AdS_5$ metric $G^{(0)}_{MN}$ in the presence
of the classical string in $AdS_5$ space, corresponding to either the 
trailing string or the accelerating string. We are only dealing with
the energy density, i.e., the $(0,0)-$component of $T_{\mu\nu}$. It
can be obtained from $h_{00}$, one of the scalar modes in the Fourier
decomposition of metric perturbations in position space $R^3$.

\subsection{Energy density for the straight string with constant velocity $v$ in the vacuum}

\begin{figure}
\begin{center}
\includegraphics[width=10cm]{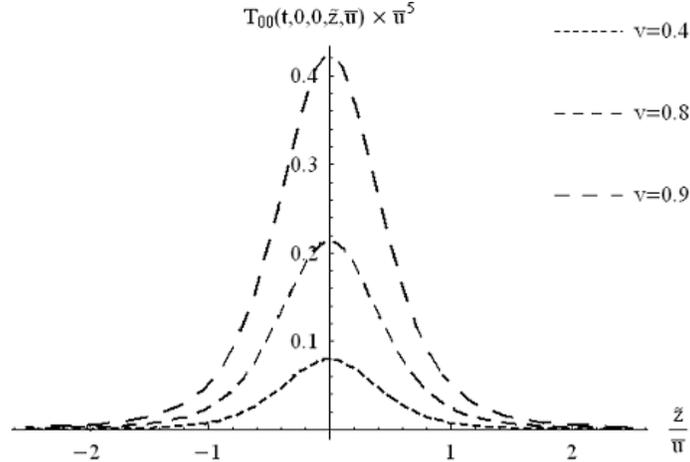}
\end{center}
\caption{$T_{00}(t,0,0,\tilde{z},\bar{u})$ of the straight string with
velocity $v=0.4,0.8,~\mbox{and}~0.9$ as a function of $\tilde{z}/\bar{u}$. $T_{00}$ peaks at $\tilde{z}=0$ and has a width $\Delta\tilde{z}\simeq\bar{u}$, which implies that a portion of straight string around $\bar{u}$ can only be projected predominantly onto the
energy density of the quark with $|\tilde{z}|\leq \bar{u}$.}\label{energydensity1}
\label{tmunu}
\end{figure}

As argued in Sec. II, our criteria to distinguish between the quark
part and the radiated energy part of a (trailing or accelerating)
string is to compare it with a straight string moving with the same velocity
$v$ in the vacuum and see which part is like that of a straight
string and which part is lagging far behind it. To see how a portion of a straight string
around $\bar{u}\equiv{1\over u}$ with velocity $v$ is related to its
energy density in position space, we define the energy density in
$(\vec{x},\bar{u})$ space for the straight string as \footnote{Here, we
use the equation of motion for the scalar field given in Ref.
\cite{Chesler:2007sv}. Note that our definition of $T_{00}(t,\vec{x},\bar{u})$ is not
quite such that $\int d\bar{u}T_{00}(t,\vec{x},\bar{u})=T_{00}(t,\vec{x}),$ but its
purpose is rather to explain our picture of Section II.
In the accelerating string calculation,
$T^{as}_{00}(t,\vec{x},\bar{u})$ will be defined from the master
equation of Ref. \cite{Gubser:2007nd}.}

\begin{equation}
T^{ss}_{00}(t,\vec{x},\bar{u})\equiv-{R^3 \over 6 \kappa_5^2}\int {
d^4 k \over (2\pi)^4 } k^2 {S(k) \over Q^2}
K_2(Q\bar{u}) e^{ikx}, \label{sssource}
\end{equation}
where $Q=\sqrt{k^2-\omega^2 }$ and the source
term\cite{Chesler:2007sv}
\begin{equation}
S(k)=-{\kappa_5^2 \sqrt{\lambda} \over R^3 }
{[k_\bot^2(2+v^2)+2(1-v^2)k_3^2]Q^2 \over k^2
\sqrt{1-v^2}}\delta(\omega-vk_3).
\end{equation}
Inserting $S(k)$ into (\ref{sssource}), we get
\begin{equation}
T^{ss}_{00}(t,\vec{x},\bar{u})={\sqrt{\lambda } \over 8\pi ^2
(1-v^2)}\left[ \frac{5 \bar{u}^2}{(\bar{u}^2+\tilde{x}^2)^{7/2}}+ {
v^2 \over 6 } \frac{10 \bar{u}^4-2 \tilde{x}^4+6 \tilde{x}^2
\tilde{z}^2-7 \bar{u}^2 (\tilde{x}^2-3 \tilde{z}^2)} {\bar{u}^2
(\bar{u}^2+\tilde{x}^2)^{7/2}}\right],\label{tss00u}
\end{equation}
where $\tilde{z}\equiv{z - vt \over \sqrt{1-v^2}}$ and
$\tilde{x}=\sqrt{x^2+y^2+\tilde{z}^2}$.

Notice that, as shown in Fig.\ref{tmunu}, a portion of straight
string around $\bar{u}$ can only be projected predominantly onto the
energy density of the quark with $|\tilde{z}|\leq \bar{u},$ that is
$|z(\bar{u})-vt|\leq {\bar{u}\over \cosh\eta }$. This is just the
infrared-ultraviolet correspondence. Also, after adding missing terms
\cite{Chesler:2007sv} and integrating out $\bar{u},$ we have
\begin{equation}
T^{ss}_{00}(t,\vec{x})={\sqrt{\lambda } \over 12 \pi ^2 (1-v^2)}\frac{(1+v^2) \tilde{x}^2-2 v^2 \tilde{z}^2 }{ \tilde{x}^6}.
\end{equation}

\subsection{Saturation momentum from the four-dimensional energy-momentum
tensor of the trailing string}
For the trailing string, the equation of motion of the scalar mode
is much more complicated\cite{Chesler:2007sv,Gubser:2007nd}, and we
will only keep terms up to $\mathcal{O}(T^2)$. In this
approximation, we can evaluate $T^{ts}_{00}(t,\vec{x},\bar{u})$ as

\begin{equation}
\begin{split}
T^{ts}_{00}(t,\vec{x},\bar{u})\equiv&-{R^3 \over 6 \kappa_5^2}\int {
d^4 k \over (2 \pi)^4 } k^2 {S(k) \over Q^2}
K_2(Q\bar{u}) \exp\{-i\omega (t-{1\over 3}(\pi
T)^2\bar{u}^3)+ik_{\perp}\cdot x_{\perp}+ik_3z\}\\
=&{\sqrt{\lambda} \over 12\pi\sqrt{1-v^2}}\int { d^3 k \over
(2\pi)^3 } [k_\bot^2(2+v^2)+2(1-v^2)k_3^2]K_2(Q\bar{u})
\exp\{ik_{\perp}\cdot
x_{\perp}+ik_3\bar{z}\}\\
=&{\sqrt{\lambda } \over 8\pi ^2 (1-v^2)}\left[ \frac{5
\bar{u}^2}{(\bar{u}^2+\tilde{x}^2)^{7/2}}+ { v^2 \over 6 } \frac{10
\bar{u}^4-2 \tilde{x}^4+6 \tilde{x}^2 \tilde{z}^2-7 \bar{u}^2
(\tilde{x}^2-3 \tilde{z}^2)} {\bar{u}^2
(\bar{u}^2+\tilde{x}^2)^{7/2}}\right].
\end{split}
\end{equation}
where $\bar{z}\equiv z - vt+{v\over 3}(\pi T)^2\bar{u}^3$,
$\tilde{z}\equiv{\bar{z}\over \sqrt{1-v^2}}$ and
$\tilde{x}=\sqrt{x^2+y^2+\tilde{z}^2}$. Comparing the expressions
for $T_{00}(t,\vec{x},\bar{u})$ of the straight string and trailing
string, the only difference is in the definition of $\tilde{z}$ in
the two cases. If ${v\over 3}(\pi T)^2\bar{u}^3\ll {\bar{u}\over
\cosh\eta}$, there is essentially no difference between the straight
and trailing strings in the region $|z-vt|\leq
{\bar{u}\over \cosh\eta}$ where almost all of the energy of the
straight string is located. If ${v\over 3}(\pi T)^2\bar{u}^3\gg
{\bar{u}\over \cosh\eta}$, the contribution to the energy of the
trailing string is much more spread out in $z-vt$ than for the
straight string and we conclude that this part of the trailing
string is not actually a part of the heavy quark but represents energy
radiated in the medium. We note that the transition point between
parts of the trailing string that belong the heavy quark and those
that are waves in the plasma is at $(\pi T)^2\bar{u}^3\sim
{\bar{u}\over \cosh\eta}$ or $u\sim \pi T \sqrt{\cosh\eta}$ which is
the natural separation point discussed in Sec. II.
\par
This separation point can also be obtained when looking at the $\bar{u}-$integrated
$T_{00},$ in the rest frame of the heavy quark: 
Ref. \cite{Yarom:2007ni} gives all the components of $T_{\mu\nu}$
(up to $\mathcal{O}(T^2)$) in the plasma rest frame, and boosting to the quark rest
frame gives
\begin{equation}
\bar{T}^{00}_{ts}\frac{\sqrt{\lambda}}{12\pi^2x^4}
-\frac{5\sqrt{\lambda}}{72}T^2\cosh{\eta} \frac{vz}{x^{3}}+
\mathcal{O}(T^4),
\end{equation}
where the contribution of the plasma alone has been subtracted. 
Equating the two terms, one sees that the quark field is unchanged
when $(\pi T)^2\cosh{\eta} v z|x|\lesssim 1.$ On the longitudinal axis ($x=y=0$), this means
$Q_sz\lesssim 1$ with $ Q_s=\pi T\sqrt{\cosh\eta}$. Or more
generally, so long as $|\vec{x}|\leq 1/Q_s$ the energy density
agrees well with that of a heavy quark in the vacuum while when
$|\vec{x}|\geq 1/Q_s$ there are strong medium modifications.

\subsection{Energy density for an accelerating string in the vacuum}
For the accelerating string $x^2=t^2+b^2-{1 \over u^2}$, the
classical stress energy tensor in momentum space is (with $M,N=(t,x,y,z,u/R^2)$)
\begin{equation}
\begin{array}{ll}
    t_{MN}(t,\vec{k},u) = {\kappa_5^2 \over \pi \alpha'} \Theta\left( u - { 1 \over \sqrt{ t^2 - x^2 + b^2 }} \right ) { 1 \over R u
    }\\
    \\
   \times \left(
        \begin{array}{ccccc}
            { b^2 + t^2 \over b x } \cos(k_1 x)  & i { t \over b } \sin(k_1 x) & 0 & 0 & {1 \over (Ru)^3} {R t \over b x} \cos(k_1 x) \\
            i { t \over b } \sin(k_1 x) & {1 \over u^2 } { t^2 u^2 -1 \over b x } \cos(k_1 x)
            & 0 & 0 & i {1 \over (Ru)^3} {R \over b} \sin(k_1 x)\\
            0 & 0 & 0 & 0 & 0 \\
            0 & 0 & 0 & 0 & 0 \\
            {1 \over (Ru)^3} {R t \over b x} \cos(k_1 x) & i {1 \over (Ru)^3} {R \over b} \sin(k_1
            x) & 0 & 0 & { 1 \over R^4 u^6 } { 1 - b^2 u^2 \over b x
            } \cos(k_1 x)
        \end{array}
        \right),\label{tmnas}
        \end{array}
\end{equation}
where $x=\sqrt{ t^2 + b^2 - {1 \over u^2}}$. To simplify our
equations below, we will use $\bar{u}=1/u$ in the following
calculation. In the zero temperature case, the master equation for
the scalar modes of the metric perturbation due to the presence of
classical string solutions is \cite{Gubser:2007nd}
\begin{equation}
\varphi _s^{''}(\bar{u})+ {1 \over \bar{u}}\varphi_s^{'}(\bar{u})-(
k^2-\omega^2 )\varphi_s(\bar{u})=-J_s(\bar{u}),
\label{scalareq}\end{equation}
where the scalar master field $\varphi _s(\bar{u}) = {P_s \over \bar{u}
} + Q_s^{tot} + ...  $, which gives the energy density as
$T_{00}(t,\vec{x})=\int { d^4 k \over (2\pi)^4 } e^{ikx} Q_s^{tot} $
and the source term 
\begin{equation}
J_s(\bar{u})= \int dt e^{i\omega t} S(\bar{u},t),
\end{equation}
is a linear combination of $t_{MN}$\cite{Gubser:2007nd}, and
$S(\bar{u},t)$ for the accelerating string is as follows,
\begin{equation}
\begin{split}
S(\bar{u},t)\equiv&-\frac{\sqrt{\lambda }}{6 b k^2 \bar{u}^5 \omega  x^3}\\
&\left\{ (k_1 \sin(k_1 x) x (2 k^2 \bar{u}^2 (i t (-3+k^2
\bar{u}^2)+3 \bar{u}^2 \omega ) x^2+6
(k^2-3 k_1^2) R^4 (t^2-\bar{u}^2) \omega  x^{'2}+\right.\\
&R^2 x (\bar{u} (2 i k^4 t \bar{u}^2-3 (k^2-3 k_1^2) (3
t^2+\bar{u}^2) \omega ) x^{'}-3 (k^2-3 k_1^2) R^2 (t^2-\bar{u}^2)
\omega  x^{''}))+\\
&\cos(k_1 x) (-6 i k^2 k_1^2 R^2 t \bar{u} x^3 x^{'}+6 (k^2-3 k_1^2)
R^4
(t^2-\bar{u}^2) \omega  x^{'2}+\\
&x^2 (\bar{u}^2 (2 k^4 \bar{u}^2 (2 i t-(b^2+2 t^2-3 \bar{u}^2)
\omega )-9 k_1^2 (t^2-\bar{u}^2) \omega  (1+\bar{u}^2 \omega ^2)+\\
&3k^2 (t^2-\bar{u}^2) \omega
 (1+\bar{u}^2 (4 k_1^2+\omega ^2)))-
3 k_1^2 (k^2-3 k_1^2) R^4 (t^2-\bar{u}^2) \omega  x^{'2})+\\
&\left.R^2 x (\bar{u} (2 i k^4 t \bar{u}^2-3 (k^2-3 k_1^2) (3
t^2+\bar{u}^2) \omega ) x^{'}-3 (k^2-3 k_1^2) R^2 (t^2-\bar{u}^2)
\omega x^{''})))\right\} \\
=&S_q(\bar{u},t)e^{-ik_1x_{as}(\bar{u},t)}+S_{\bar{q}}(\bar{u},t)e^{ik_1x_{as}(\bar{u},t)},
\end{split}
\end{equation}
with $x=x_{as}(\bar{u},t)=\sqrt{t^2+b^2-\bar{u}^2}$.
From (\ref{scalareq}), it is easy to show that
\begin{equation}
Q_s^{tot}=\lim\limits_{\epsilon\rightarrow 0} \{\int_\epsilon^\infty
d\bar{u} \bar{u} K_0(Q\bar{u})J_s(\bar{u})- \mbox{ poles at
} 0\}.
\end{equation}
To compare an accelerating string with a straight string, we write
\begin{equation}
\begin{split}
T^{as}_{00}(t,\vec{x},u)=&\int { d^4 kd\tilde{t} \over (2\pi)^4 }
\bar{u}
K_0(Q\bar{u})S(\bar{u},\tilde{t}) e^{ikx}\\
=&\int { d^4 k d\tilde{t} \over (2\pi)^4 } \bar{u}
K_0(Q\bar{u})\\
&\times\left\{S_q(\bar{u},\tilde{t})\exp[-i\omega(t-\tilde{t})+ik_1(x-x_{as}(\bar{u},\tilde{t}))+ik_{\perp}
\cdot
x_{\perp}]\right.\\
&\left.+S_{\bar{q}}(\bar{u},\tilde{t})\exp[-i\omega(t-\tilde{t})+ik_1(x+x_{as}(\bar{u},\tilde{t}))+ik_{\perp}
\cdot x_{\perp}]\right\},\label{energydensityinux}
\end{split}
\end{equation}
or, in momentum space,

\begin{equation}
T^{as}_{00}(\omega,\vec{k},u)=\bar{u}
K_0(Q\bar{u})J_s(\bar{u}).\label{energydensityinuk}
\end{equation}
The Bessel $K$ functions and Fourier transformations between
position space and momentum space ensure the infrared-ultraviolet
correspondence. We will evaluate $T^{as}_{00}(t,\vec{x},u)$ or
$T^{as}_{00}(\omega,\vec{k},u)$ in the approximation $t>>b,\bar{u}$
and only keep terms of order $\mathcal{O}({t \over b})$. In
(\ref{energydensityinux}), we integrate over the region
$t-\mathcal{T}<\tilde{t}<t+\mathcal{T}$ where we choose a
$\mathcal{T}$ which is much smaller than t but large enough to
ensure we can use the $\delta$ function as a good approximation to
the resulting $\tilde{t}$ integration. In our calculation we
consider the contribution to $T_{00}(t,\vec{x},\bar{u})$ of an
element of matter on the string, located at $\bar{u}$ at time $t$,
and moving along the string at velocity
$v(\bar{u},t)={x_{as}(\bar{u},t) \over t^2+b^2}\simeq 1-{\bar{u}^2
\over 2t^2}$ as given in (\ref{velocity}) of Sec. V. In case (i)
below where $\bar{u}$ is small it will not be necessary to
distinguish between $\dot{x}_{as}(\bar{u},t)={\partial\over\partial
t}x_{as}(\bar{u},t)$ and $v(\bar{u},t)$ since
${\partial\over\partial \bar{u}}x_{as}(\bar{u},t)$ is so small that
the difference of these two velocities is unimportant. In case (ii)
below where $\bar{u}$ is relatively large this distinction is
important, and the natural organization of the calculation is to
evaluate the contribution to $T_{00}$ of a fixed element of matter,
following its motion on the string, over a period of time
2$\mathcal{T}$. Of course this procedure is effective only because
the motion of matter on the string between time $t-\mathcal{T}$ and
$t+\mathcal{T}$ is small,
$\bar{u}(t+\mathcal{T})-\bar{u}(t-\mathcal{T})\simeq
{\partial\bar{u}\over\partial
t}2\mathcal{T}
\simeq\bar{u}{2\mathcal{T}\over t}\ll\bar{u}$ when
${2\mathcal{T}\over t}\ll 1$. In this large $t$ approximation, we
have
\begin{equation}
\begin{split}
T^{as}_{00}(t,\vec{x},u)=&\int { d^4 k \over (2\pi)^4 }
\bar{u}\int^{t+\mathcal{T}}_{t-\mathcal{T}}d\tilde{t}
K_0(Q\bar{u})\\
&\times\{S_q(\bar{u},\tilde{t})\exp[-i\omega(t-\tilde{t})+ik_1(x-x_{as}(\bar{u},\tilde{t}))+ik_{\perp}
\cdot
x_{\perp}]\\
&+S_{\bar{q}}(\bar{u},\tilde{t})\exp[-i\omega(t-\tilde{t})+ik_1(x+x_{as}(\bar{u},\tilde{t}))+ik_{\perp}
\cdot x_{\perp}]\}\\
\simeq& \int { d^4 k \over (2\pi)^4 }
\bar{u}\int^{\mathcal{T}}_{-\mathcal{T}}d\bar{t}
K_0(Q\bar{u})\\
&\times\{S_q(\bar{u},t)\exp[i\omega
\bar{t}+ik_1(x-x_{as}(\bar{u},t)-v(\bar{u},t)\bar{t})+ik_{\perp}
\cdot
x_{\perp}]\\
&+S_{\bar{q}}(\bar{u},t)\exp[i\omega
\bar{t}+ik_1(x+x_{as}(\bar{u},t)+v(\bar{u},t)\bar{t})+ik_{\perp}
\cdot x_{\perp}]\}\\
\simeq& \int { d^4 k \over (2\pi)^3 } \bar{u} K_0(Q\bar{u})\\
&\times\{\delta(\omega-k_1v(\bar{u},t))S_q(\bar{u},t)\exp[ik_1(x-x_{as}(\bar{u},t))+ik_{\perp}
\cdot
x_{\perp}]\\
&+\delta(\omega+k_1v(\bar{u},t))S_{\bar{q}}(\bar{u},t)\exp[ik_1(x+x_{as}(\bar{u},t))+ik_{\perp}
\cdot x_{\perp}]\}
\end{split}
\end{equation}
Therefore, in the end, we arrive at
\begin{equation}
T^{as}_{00}(t,\vec{x},u) \simeq \int { d^4 k\over (2\pi)^4 } \bar{u}
K_0(Q\bar{u})J_s(\bar{u}),
\end{equation}

We will compare the following two cases:

(i) $t\gg b\gg \bar{u}$,

In our picture, the two parts of the accelerating string are the
parts of a quark and an anti-quark respectively. We can show that in
the large $t$ limit, the energy density related to these two parts
of the string is simply a sum of that of a quark and an anti-quark.
In this case, $v(\bar{u},t)\simeq 1$ and the source term
$J_s(\bar{u})$ is

\begin{equation}
\begin{split}
J_s(\bar{u})\simeq&{\sqrt{\lambda } t \over b }\left\{\delta (\omega
- k_1) \left[-\frac{2 k_1^2 }{\bar{u}}-\frac{\omega ^2 }{2
\bar{u}}+\frac{2 k^2 }{3 \bar{u}}+\frac{ k_1k^2 }{3 \bar{u} \omega
}+\frac{3k_1^2 \omega ^2 }{2 k^2 \bar{u}}-\frac{1}{2
\bar{u}^3}+\frac{3 k_1^2 }{2 k^2 \bar{u}^3}-\frac{k_1}{\bar{u}^3
\omega }\right]\right.\\
&+\left.\delta (\omega + k_1) \left[-\frac{2 k_1^2
}{\bar{u}}-\frac{\omega ^2 }{2 \bar{u}}+\frac{2 k^2 }{3
\bar{u}}-\frac{ k_1k^2 }{3 \bar{u} \omega }+\frac{3k_1^2 \omega ^2
}{2 k^2 \bar{u}}-\frac{1}{2 \bar{u}^3}+\frac{3 k_1^2 }{2 k^2
\bar{u}^3}+\frac{k_1}{\bar{u}^3 \omega }\right]\right\},
\end{split}
\end{equation}
where ${t\over b}\simeq \cosh\eta_0$ in the large t limit. Compared
with the source term of a straight string with velocity
$v$\cite{Gubser:2007nd},
\begin{equation}
\begin{split}
J_s^{ss}(\bar{u})&={\sqrt{\lambda }\delta(\omega-vk_1) \over
\sqrt{1-v^2} }\left[-\frac{2 k_1^2 v^2 }{\bar{u}}-\frac{v^2 \omega
^2}{2 \bar{u}}+\frac{(1+v^2)k^2 }{3 \bar{u}}+\frac{k^2 k_1 v }{3
\bar{u} \omega }+\frac{3 k_1^2 v^2
 \omega ^2}{2 k^2 \bar{u}}-\frac{v^2}{2
\bar{u}^3}+\frac{3 k_1^2 v^2 }{2 k^2 \bar{u}^3}-\frac{k_1 v
}{\bar{u}^3 \omega }\right]\\
&\stackrel{v\rightarrow 1}{\longrightarrow}{\sqrt{\lambda
}\delta(\omega-k_1) \over \sqrt{1-v^2} }\left[-\frac{2
k_1^2}{\bar{u}}-\frac{ \omega ^2}{2 \bar{u}}+\frac{2k^2 }{3
\bar{u}}+\frac{k^2 k_1}{3 \bar{u} \omega }+\frac{3 k_1^2
 \omega ^2}{2 k^2 \bar{u}}-\frac{1}{2
\bar{u}^3}+\frac{3 k_1^2 }{2 k^2 \bar{u}^3}-\frac{k_1 }{\bar{u}^3
\omega }\right],
\end{split}
\end{equation}
so that, $J_s(\bar{u})$ is simply a sum of that of a quark and an
anti-quark flying apart with constant velocity $v$ close to 1. Thus
in the region $\bar{u}\ll b$ and for $t\gg b$ the energy density of
an accelerating string is indistinguishable from that of a string
moving at constant velocity with ${t\over b}=\cosh\eta_0$.

(ii)$t\gg\bar{u} \gg b$,

In our picture, this part of string corresponds to radiated energy
and we will see that it contributes predominantly to the energy
density far behind the quark. Recalling that in this case
$v(\bar{u},t)={x_{as}(\bar{u},t) \over t^2+b^2}\simeq 1-{\bar{u}^2
\over 2t^2}$, and $\cosh\eta\simeq{t\over \bar{u}}$, let us focus
only on the half string associated with a quark, where the energy
density at $(\vec{x},\bar{u})$ is
\begin{equation}
\begin{split}
T^{as}_{00}(t,x,0_{\perp},u)=&\int { d^4 k\over (2\pi)^4 }
\exp\{ik_1[x-x_{as}(\bar{u},t)]\}\bar{u}
K_0(Q\bar{u})J_s(\bar{u})\}\\
\simeq&{\sqrt{\lambda } t \over b }\int { d^4 k\over (2\pi)^4 }
\bar{u}K_0(Q\bar{u})\exp\{ik_1[x-x_{as}(\bar{u},t)]\}\\
&\times\left\{\delta (\omega - k_1(1-{\bar{u}^2\over 2t^2}))
\left[-\frac{2 k_1^2 }{\bar{u}}-\frac{\omega ^2 }{2 \bar{u}}+\frac{2
k^2 }{3 \bar{u}}+\frac{ k_1k^2 }{3 \bar{u} \omega }\right.\right.\\
&\left.\left.+\frac{3k_1^2 \omega ^2 }{2 k^2 \bar{u}}-\frac{1}{2
\bar{u}^3}+\frac{3 k_1^2 }{2
k^2 \bar{u}^3}-\frac{k_1}{\bar{u}^3 \omega }+\mathcal{O}({u \over t})\right]\right\}\\
\simeq&{\sqrt{\lambda } t \over b }\int { d
k_1dk_{\perp}k_{\perp}\over (2\pi)^3 }
K_0(\tilde{k}\bar{u})\exp\{ik_1[x-x_{as}(\bar{u},t)]\}\\
&\times \left[-\frac{5 k_1^2 }{2}+k^2 +\frac{3k_1^4 }{2
k^2}-\frac{3}{2 \bar{u}^2}+\frac{3 k_1^2 }{2 k^2
\bar{u}^2}+\mathcal{O}({u \over t})\right],
\end{split}
\end{equation}
where $\tilde{k}\equiv\sqrt{{\bar{u}^2\over t^2}k_1^2+k_{\perp}^2
}$. The exponential decrease of the Bessel $K_0$ function at large
values of its argument requires ${\bar{u}\over t} k_1,
k_{\perp}\sim 1/\bar{u}$, which means ${k_1^2 \over k^2}\simeq 1$
and ${k_{\perp}^2 \over k_1^2}\ll 1$, so in this approximation, we
have

\begin{equation}
\begin{split}
T^{as}_{00}(t,x,0_{\perp},u)\simeq& {\sqrt{\lambda } t \over b }\int
{ d k_1dk_{\perp}k^3_{\perp}\over (2\pi)^3 }
K_0\left(\tilde{k}\bar{u}\right)\exp\left\{ik_1[x-x_{as}(\bar{u},t)]\right\}\\
=&{3\sqrt{\lambda }\over 4 \pi^2 }{t \over b }{t\over \bar{u}}{1
\over {(\bar{u}^2 + {t^2\over u^2}\bar{x}^2)^{5/2}}},
\end{split}
\end{equation}
where $\bar{x}\equiv x-x_{as}(\bar{u},t)$. The dominant term of
$T^{as}_{00}(t,x,0_{\perp},\bar{u})$ of order $\mathcal{O}({t\over
b})$ shows that this part of the accelerating string indeed
contributes predominantly around $x_{as}(\bar{u},t)$ with
$\Delta(x-x_{as}(\bar{u},t))\simeq {\bar{u}^2 \over t}={\bar{u}
\over \cosh\eta}$ while $x_{as}(0,t)-x_{as}(\bar{u},t)\simeq
{\bar{u}^2 \over 2t}$. Although there is some contribution which
could be associated with the energy density of the quark, it is
quite small relative to the part of string with $\bar{u}\ll b$,
since it is suppressed by a factor $\cosh\eta/\cosh\eta_0={b\over
\bar{u}}\ll 1$ due to their different Lorentz contractions of the
local part of the string at $\bar{u}$ and that of the overall motion
of the heavy quark. On the other hand, the quark part of the string
with $\bar{u}\ll b$ barely contributes to this part of energy of the
"gluonic" field since its contribution is more localized around the
position of the quark due to much larger Lorentz contraction
$\cosh\eta_0={t \over b}$.
\subsection{Seeing the separation point on the trailing string from causality }
There is a simple argument, although not directly related to the
energy-momentum tensor, which shows that the point $u=\pi T
\sqrt{\cosh\eta}=Q_s$ separates the trailing string into a part
which is causally connected from a part which is not causally
connected to the core of the heavy quark. In $AdS_5$ space, we can
boost the plasma rest frame into the quark and the trailing string
rest frame by the following generalized Lorentz transformation,
\begin{eqnarray}
\hat{x}&=&\cosh\eta\left[ x - v ( t + F(u) )\right],\\
\hat{t}&=&\cosh\eta\left[ t - v x + F(u) - \cosh^{-{3\over
2}}\eta F\left({u\over \sqrt{\cosh\eta}}\right) \right],\\
\hat{y}&=&y,\hat{z}=z,\hat{u}=u,
\end{eqnarray}
where $F(u)=\frac{1}{2u_{h}}\left[ \frac{\pi }{2}-\tan ^{-1}\left(
\frac{u}{u_{h}}\right) -\coth ^{-1}\left( \frac{u}{u_{h}}\right)
\right]$ and we take $z_0=0$ in (\ref{shape}). In this frame, the
trailing string appears as a straight string sitting at $\hat{x}=0$
at all time $\hat{t}$, and the $AdS_5$ metric is
\begin{equation}
\begin{split}
ds^2=&R^2u^2\left[-\hat{f}(u)d\hat{t}^2+2v\cosh^2\eta{u^4_h\over
u^4}d\hat{x}d\hat{t}+(1+v^2\cosh^2\eta{u^4_h\over
u^4})d\hat{x}^2+dy^2+dz^2\right]\\
&+{2R^2v\cosh\eta u^2_hd\hat{x}du\over u^2\hat{f}(u)}+{du^2R^2\over
u^2\hat{f}(u)},
\end{split}
\end{equation}
where $\hat{f}(u)\equiv 1-{u^4_h\cosh^2\eta\over u^4}$. From this
metric, it is easy to see that it takes a beam of light emitted at
$u=u_h\sqrt{\cosh\eta}=\pi T\sqrt{\cosh\eta}$, the separation point
that we have discussed earlier, an infinite period of time $\Delta
\hat{t}=\Delta t/\cosh\eta$ to reach the quark, that is, the quark
can never "see" the part of the string with $u\leq\pi
T\sqrt{\cosh\eta}$.


\end{document}